\documentclass[review]{elsarticle}

\usepackage{graphicx}
\usepackage{nicefrac}
\usepackage{amsmath}
\usepackage{amssymb}
\usepackage{hyperref}

\journal{Physica A}

\begin{document}

	\begin{frontmatter}
		\title{Relevant Stylized Facts About Bitcoin: Fluctuations, First Return Probability, and Natural Phenomena}
		
		\author{C. R. da Cunha}
		\ead{creq@if.ufrgs.br}
		\author{R. da Silva}
		\address{Instituto de F\'isica, Universidade Federal do Rio Grande do Sul, 91501-970 Porto Alegre, Rio Grande do Sul, Brazil}

		\begin{abstract}
			Bitcoin is a digital financial asset that is devoid of a central authority. This makes it distinct from traditional financial assets in a number of ways. For instance, the total number of tokens is limited and it has not explicit use value. Nonetheless, little is know whether it obeys the same stylized facts found in traditional financial assets. Here we test bitcoin for a set of these stylized facts and conclude that it behaves statistically as most of other assets. For instance, it exhibits aggregational Gaussianity and fluctuation scaling. Moreover, we show by an analogy with natural occurring quakes that bitcoin obeys both the Omori and Gutenberg-Richter laws. Finally, we show that the global persistence, originally defined for spin systems, presents a power law behavior with exponent similar to that found in stock markets.
		\end{abstract}
	\end{frontmatter}
	

	\section{Introduction}
	Unlike tangible goods, digital tokens can be easily copied and distributed. This constitutes a major challenge in using digital media for financial transactions.
	Nonetheless, bitcoin (BTC) was proposed in 2008 as a peer-to-peer solution to this double spending problem. Transactions in this protocol are collected into blocks that are verified by all nodes of a network.\cite{Satoshi} Furthermore, the maximum number of available tokens in this solution is limited to approximately 12.6 billion units, which makes it an appropriate payment system.
	
	Nevertheless, unlike gold or other precious metal, bitcoin has no explicit use value. Rather, it has mostly exchange value. Also, it is not backed or controlled by force by any state. Thus its price is solely dictated by the forces of offer and demand. Also, it is not a tangible asset and thus has reduced transaction costs. Moreover, unlike stocks, bitcoin is negotiated non-stop worldwide.
	
	This innovation has naturally attracted the attention of the scientific community. For instance, the Hurst exponent has been measured for different time scales\cite{BTCHurst}. Its multifractality degrees are known to be higher than those of many other indices\cite{BTCMulti}. Furthermore, estimation of a bubble process has been performed on BTC \cite{BTCBubble}. 
	
	Nonetheless, little is known about its statistical nature. For instance, traditional financial assets are known to present certain regularities and general tendencies that are known as \emph{stylized facts}.\cite{SF1,SF2,EmpFacts} Does bitcoin exhibit the same stylized facts? How does it compare to those found in standard financial assets? These are central questions that we will try to answer along this paper.
	
	We will proceed by discussing the distribution of returns and its moments. Then we will present some correlations such as that between volume and volatility. We will then close the discussion making a comparison between the volatility of bitcoin and natural occurring  quakes. Furthermore, we studied phenomena related to the first return probability using the persistence of bitcoin prices as an estimator. This has originally been used to measure the probability that a spin system remains magnetized above (or below) its initial value. Nonetheless, it has been shown that the same concept can be extended to financial markets\cite{Constantin2005}.
	
	\section{Data Analyzed}
		
		The daily close prices ($p_m$) of BTC were obtained from CoinMarketCap for the period between Apr/28/2013 and Feb/14/2019. High frequency data was obtained from BitcoinCharts for the period between Jan/07/2018-00:00:00 and Feb/07/2018-11:29:00. The log returns were calculated as $r_m=\log(p_{m+1})-\log(p_m)$. Both the daily close prices and their log returns are shown in Fig. \ref{fig:Retornos}.

	\section{Results}
		
		\subsection{Probability Density}
		
		The log returns were normalized as:
		
		\begin{equation}
			n_m=\frac{r_m-\langle r\rangle}{\left(\langle r^2\rangle-\langle r\rangle^2\right)^{1/2}}.
		\end{equation}
			
		\begin{figure}
			\centering
			\includegraphics{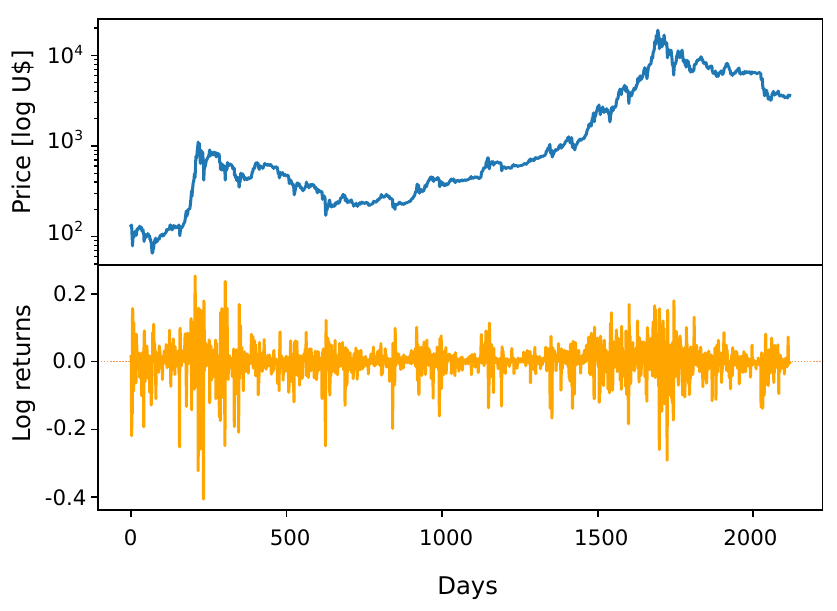}
			\caption{Top) Logarithmic of the daily bitcoin close price in US\$, and Bottom) corresponding log returns for the period between Apr/28/2013 and Feb/14/2019. Source: CoinMarketCap.}
			\label{fig:Retornos}
		\end{figure}
		
		The probability density function (PDF) was estimated with an Epanechnikov kernel and a window of size:
		
		\begin{equation}
			h = 1.06\sigma^2N^{-0.2},
		\end{equation}
		where $\sigma^2$ is the second central moment of a series with $N$ elements, $h_1\approx 1.06$, and $h_2\approx 0.2$.
		
\begin{figure}
			\centering
			\includegraphics{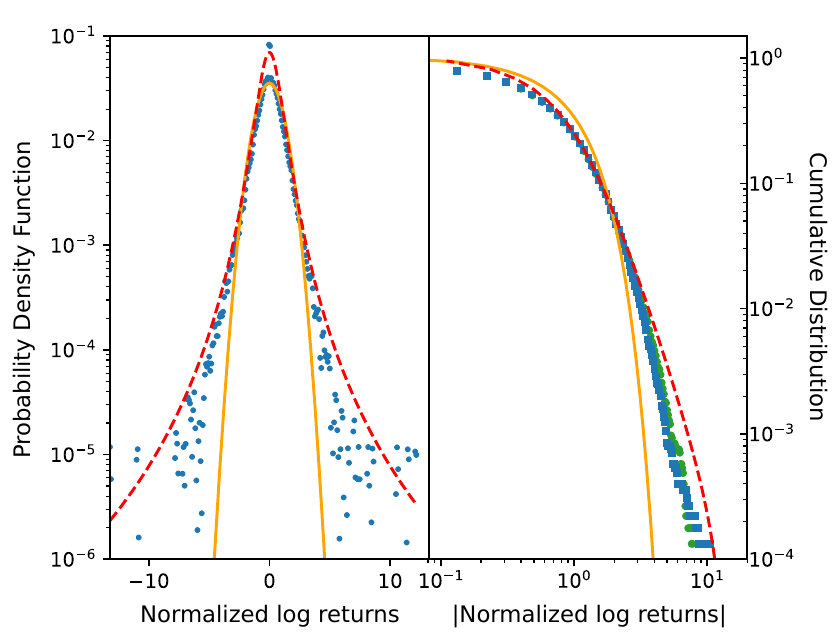}
			\caption{Left) Probability density function for the BTC minute close returns, Right) Complementary cumulative distribution for the same set of data. In both cases, dashed red curves correspond to a Student-t fitting, orange curves correspond to a Gaussian fitting, blue squares correspond to the positive tail and the green dots correspond to the negative tail.}
			\label{fig:Distribuicao}
		\end{figure}

		The complementary cumulative distribution (CCDF) was computed as $Pr(X> x)$ directly from the time series. These calculations are shown in Fig. \ref{fig:Distribuicao} together with Gaussian and Student-t fittings. As it can be visually detected, the experimental distribution has tails that are note well fitted by a Gaussian distribution. Therefore, the formation of returns does not seem to be related to a simple additive process. The CCDF decays as a power law with coefficients $\sim-2$ for the negative tail and $\sim-3$ for the positive tail.
		
		This constitutes the first stylized fact observed in financial markets and confirmed for BTC: \emph{The distribution of bitcoin returns has fat tails}.

		\subsection{Moments}
		The fat tails found in the distribution of returns impose complications to dispersion measurements. For instance, there is no guarantee that their theoretical moments are finite.
		However, it has been suggested that if the theoretical moment is finite, then the sample moment has to converge to a finite value as more data is added to the series.\cite{Mandel1} 
		
\begin{figure}
			\centering
			\includegraphics{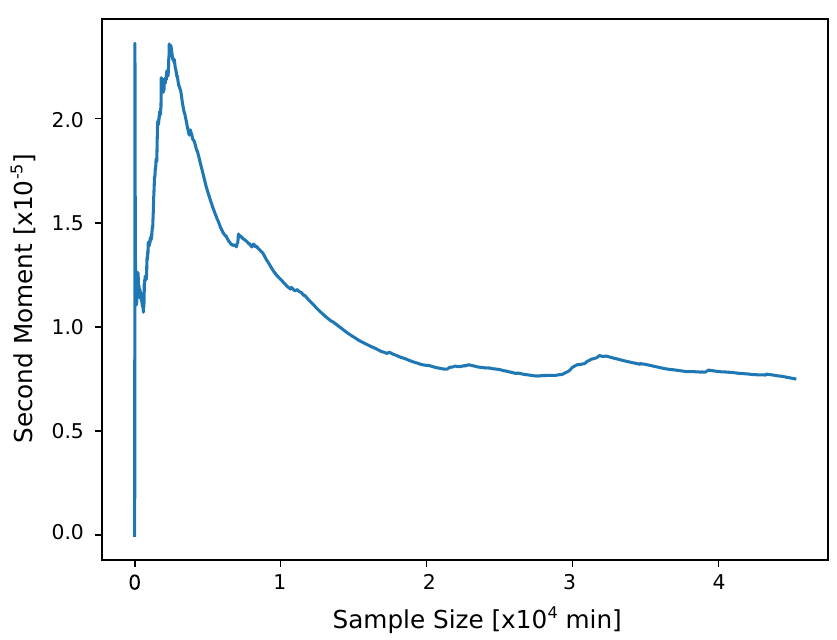}
			\caption{Second central moment as a function of the series length.}
			\label{fig:FiniteSize}
		\end{figure}

		Fig. \ref{fig:FiniteSize} shows that the second moment converges to a value around $0.7\times 10^{-5}$ ater approximately $2\times10^4$ minutes. Therefore, the theoretical variance of the distribution is finite.

		A further assessment of the distribution of returns can be analyzed by the fourth moment (kurtosis), computed as:
		
		\begin{equation}
			K(X)=\left\langle\left(\frac{X-\mu}{\sigma}\right)^4\right\rangle.
		\end{equation}
		The excess kurtosis is defined as $K(X)-3$ such that positive values indicate leptokurtic distributions, whereas negative values indicate platicurtic distributions. Since the data length influences the statistics, we used 100 bootstrap samples of 100 data points each to keep the data size constant. We used this strategy to estimate the kurtosis and further statistics that depend on the time scale.
		
\begin{figure}
			\centering
			\includegraphics{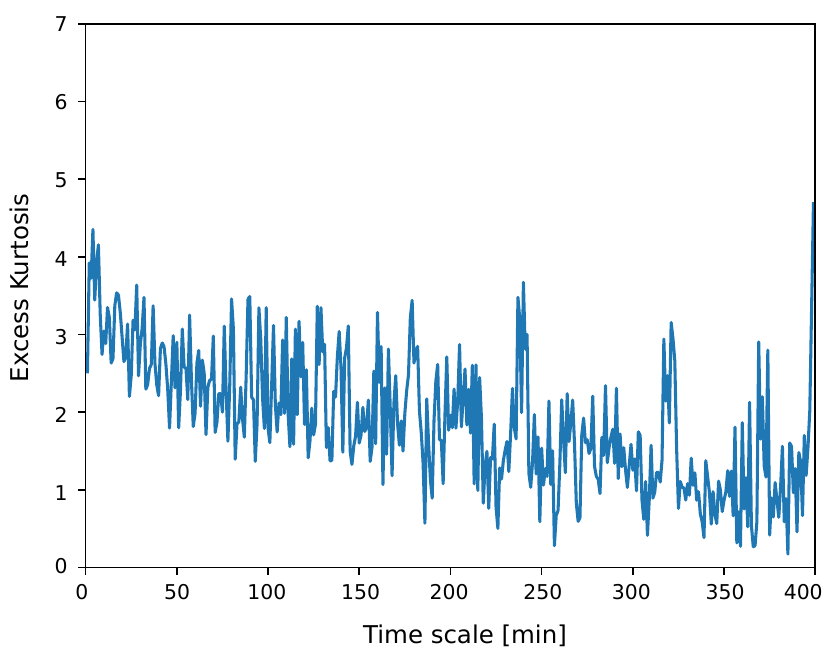}
			\caption{The excess kurtosis for different time scales. Positive values correspond to leptokurtic distributions, whereas zero corresponds to a normal distribution.}
			\label{fig:Kurt}
		\end{figure}

		The returns for different time scales were computed as:
		
		\begin{equation}
			r_{m,\tau}=\log(p_{m+\tau})-\log(p_m),
		\end{equation}
		and the kurtosis for different time scales was computed as $K(r_{m,\tau})$.		
		Figure \ref{fig:Kurt} shows the excess kurtosis for different time scales from 1 minute up to 400 minutes. 
		
		Although higher excess kurtosis values are found for other financial assets (future indexes can have $K>70$, for example), the distribution is clearly leptokurtic for short time scales. However, it tends to a Gaussian distribution as the time scale increases. This is the second verified stylized fact: \emph{Aggregational Gaussianity is observed for BTC returns.}

\begin{figure}
			\centering
			\includegraphics{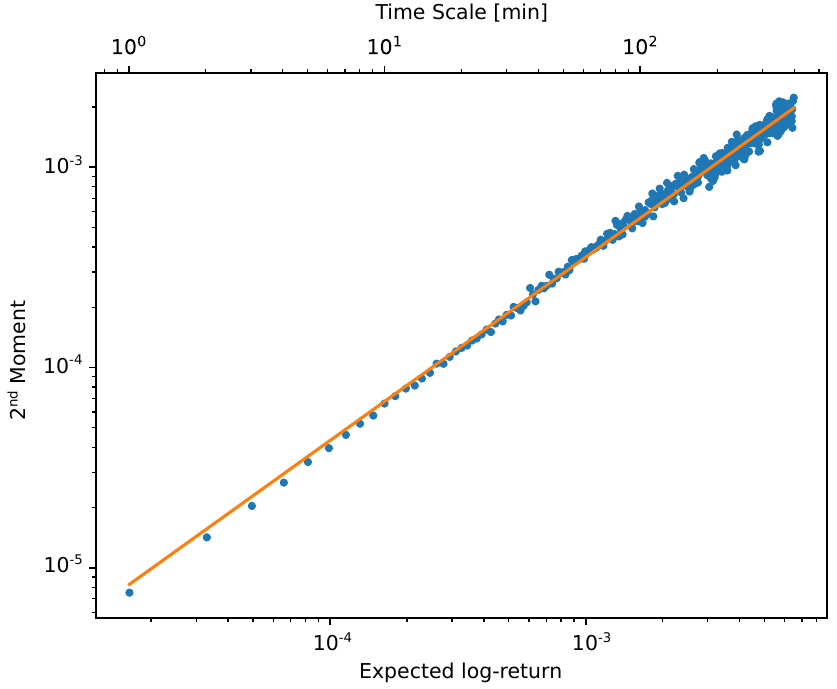}
			\caption{Second moment of the log returns as a function of the expected return for different time scales (blue dots). The solid orange line is a power-law fitting of the data. The expected return is proportional to the time scale.}
			\label{fig:secmon}
		\end{figure}

		As shown in Fig. \ref{fig:Retornos}, the price has a positive long-term trend. Therefore, the expected return increases with time scale. This makes it possible to plot the second moment as a function of the expected return as shown in Fig. \ref{fig:secmon}. The figure strongly suggests that the variance is a power-law of the expected return given as:
		
		\begin{equation}
			\text{VAR}\left\{r_{n,\tau}\right\}\sim 2\langle r_{m,\tau}\rangle^{0.916}\propto \tau^{0.912},
		\end{equation}
		where $\lambda\approx0.916\pm0.054$, and $\gamma\approx0.912$.
		This variance-to-mean power law is the signature of Taylor's law\cite{Taylor} found in many other natural systems such as in cancer metastasis\cite{Kendal} and in the human genome\cite{Kendal2}. 
		
		Although Taylor's law can be ascribed to a Tweedie distribution\cite{Tweedie}, no particular distribution is known for the parameters found in this work.	
		Nonetheless, this constitutes another stylized fact: \emph{BTC results show fluctuation scaling}.

		\subsection{Correlations}
		In this section we will analyze different correlations that appear in the time series of bitcoin returns.
		
		\subsubsection{Autocorrelation}
		
\begin{figure}
			\centering
			\includegraphics{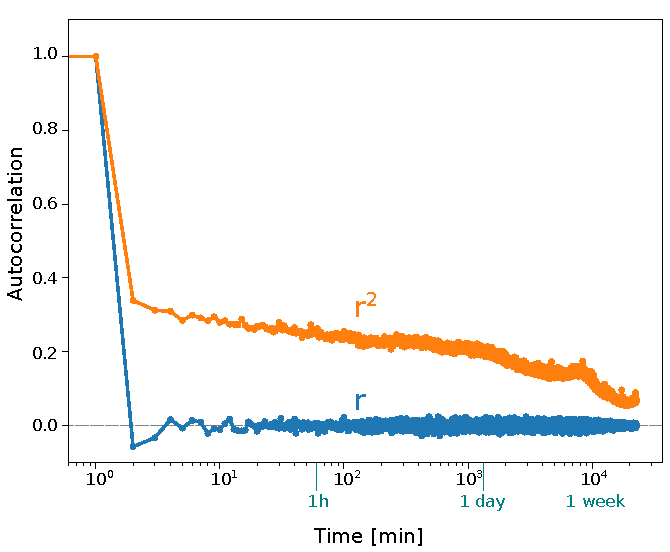}
			\caption{The autocorrelation of returns $r$ (blue) and the autocorrelation of the volatility $r^2$ (orange).}
			\label{fig:autoc}
		\end{figure}

		The autocorrelation function was calculated using the Wigner-Khinchin theorem:
		
		\begin{equation}
			A(\tau)=\mathcal{F}^{-1}\left\{R(\omega)R^*(\omega)\right\},
		\end{equation}
		where $R(\omega)$ is the Fourier transform of the returns. The autocorrelation of the returns squared was calculated the same way. The results shown in Fig. \ref{fig:autoc} indicate that, although the log returns do not show any correlation (slope in the semilog plot $=0.0113(21)$), the variance exhibits a positive persistence over several days (slope $=0.1021(26)$). Thus, periods of high volatility are followed by other periods of high volatility as well as periods of low volatility are followed by periods of low volatility.
		This is the another verified stylized fact: \emph{bitcoin exhibits volatility that tends to cluster in time}.

		\subsubsection{Volume $\times$ volatility}
		
\begin{figure}
			\centering
			\includegraphics{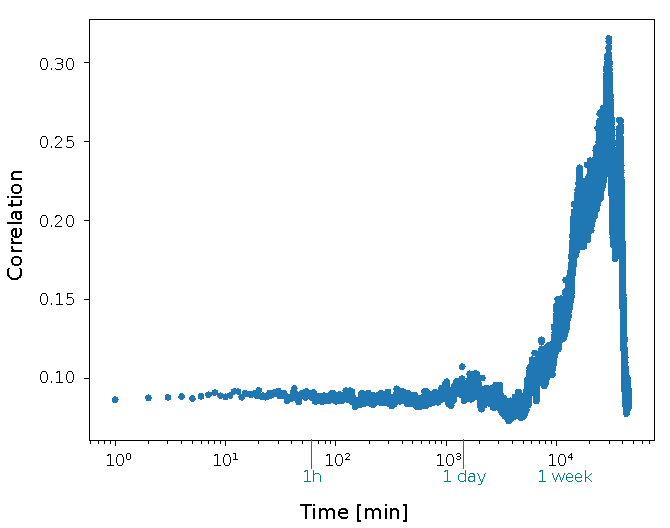}
			\caption{Correlation coefficient between the volume and the volatility.}
			\label{fig:vol2}
		\end{figure}

		The correlation function between the volume and the volatility, given by:
		
		\begin{equation}
			C_{vv}(\tau) = \left\langle \text{vol}(t+\tau)\times\sigma^2(t)\right\rangle
		\end{equation}
		was also estimated by the inverse Fourier transform and is shown in Fig. \ref{fig:vol2}. The volatility is weakly correlated to the volume in the short-term, but peaks in the medium-term. Thus, high volumes correspond to high risks, specially in the medium-term.
		This constitutes another stylized fact: \emph{The correlation between the volume and the volatility for BTC is always positive}.

		\subsubsection{Coarse Graining}
		
\begin{figure}
			\centering
			\includegraphics{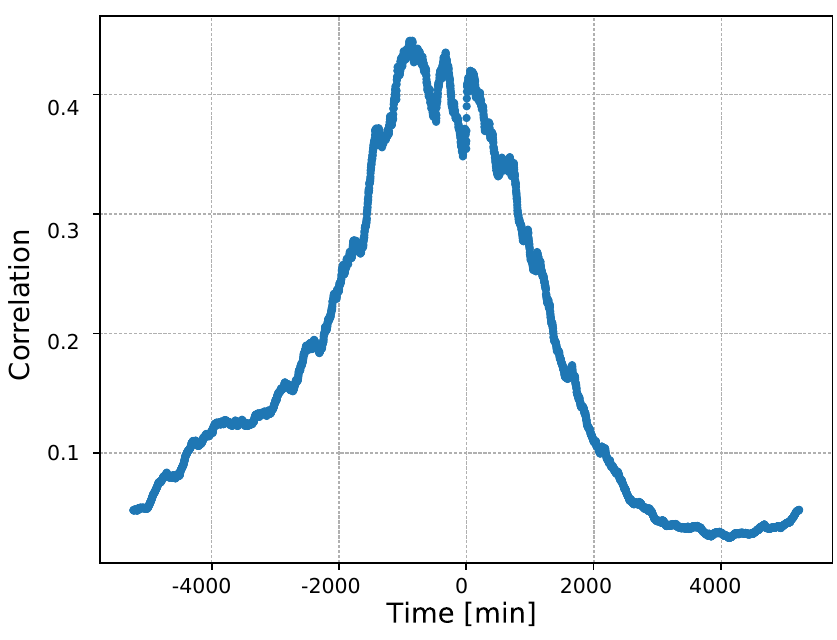}
			\caption{Correlation between the fine-scale volatility and the volatility computed for a period of 4000 time units.}
			\label{fig:ren}
		\end{figure}

		The returns were coarse grained by:
		
		\begin{equation}
			r_{cg}(t) = \log(p[t+T])-\log(p[t]),
		\end{equation}
		where $T=4000$ in our case.
		The correlation between the corresponding coarse grained volatility and the short-term volatility is given by:
		
		\begin{equation}
			C_{cg}(\tau) = \left\langle \left(r_{cg}(t+\tau)-\langle r_{cg}\rangle\right)^2 \left(r(t)-\langle r\rangle\right)^2\right\rangle.
		\end{equation}
		This computation was also done with the inverse Fourier transform and the result is shown in Fig. \ref{fig:ren}. There is a clear asymmetry between positive and negative lags, which constitute another stylized fact: \emph{The coarse-grained BTC returns predict the fine structure better than the other way around}. 

		\subsection{Financial Quakes}
		
		Seismology offers a set of tools that have been shown to work well with natural shocks. In this section we propose an analogy between financial shocks and seismic phenomena, showing that some financial stylized facts observed in BTC have the same behavior observed in these systems.
		
		Earthquakes are followed by smaller aftershocks whose frequency $n(t)$ is inversely proportional to the time elapsed after the main shock. This empirical observation,  known as Omori's law, is mathematically given by:
		
		\begin{equation}
			n(t) \propto (t-t_0)^{1-p},
		\end{equation}
		where $t_0$ is a constant corresponding to the onset of a quake, and $p$ is a constant related to the decay rate.\cite{OmoriUtsu}
		
		In order to relate financial shocks to earthquakes, an event counter was defined as:
		
		\begin{equation}
			N(t) = \sum_{t'<t} \Theta(|r(t')|-r_{th}),
		\end{equation}
		where $\Theta(t)$ is the Heaviside step function, and $r_{th}$ was chosen to be $3\sigma$.\cite{Petersen}

\begin{figure}
			\centering
			\includegraphics{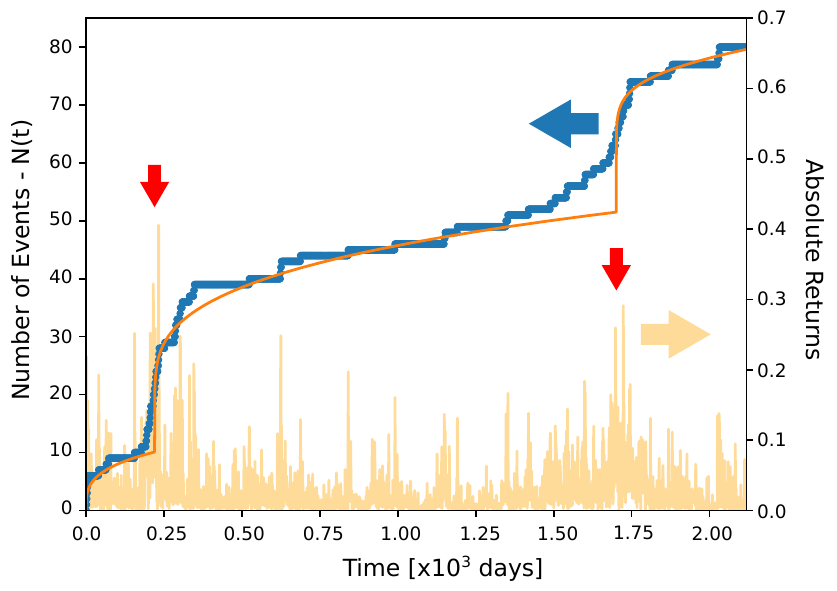}
			\caption{The weak orange line corresponds to absolute log-returns $|r(t)|$, the blue line is the events counter $N(t)$ and the solid orange curve is a coarse fitting of the events counter. Red arrows indicate the onset of abnormal returns.}
			\label{fig:Omori}
		\end{figure}

		Fig. \ref{fig:Omori} shows the empirical counter data together with a Levenberg-Marquardt fitting for a generalized Omori law of the form:
		
		\begin{equation}
			N(t) \propto \sum_{t_0} (t-t_0)^{1-p}\Theta(t-t_0),
		\end{equation}
		where, $t_0$ is indicated by red arrows in Fig. \ref{fig:Omori}. The decay rate in this case was $\sim 0.8$.

		\subsubsection{Gutenberg-Richter law}
		
		The number $N_{GR}$ of earthquakes with magnitude higher or equal to a certain threshold $M$ is given by the Gutenberg-Richter law\cite{Richter1}:
		
		\begin{equation}
			N_{GR}(M) = 10^{a-bM},
		\end{equation}
		where $a$ and $b$ are constants. 
		
		We relate the quake magnitude with the absolute logarithmic return. We support this strategy by the fact that the Richter magnitude $M_L$ has the same shape of a log-return:
		
		\begin{equation}
			M_L=\log_{10}(A)-\log_{10}A_0,
		\end{equation}
		where $A$ and $A_0$ are the excursion and a standard excursion of a seismograph.\cite{RichterScale}
		
\begin{figure}
			\centering
			\includegraphics{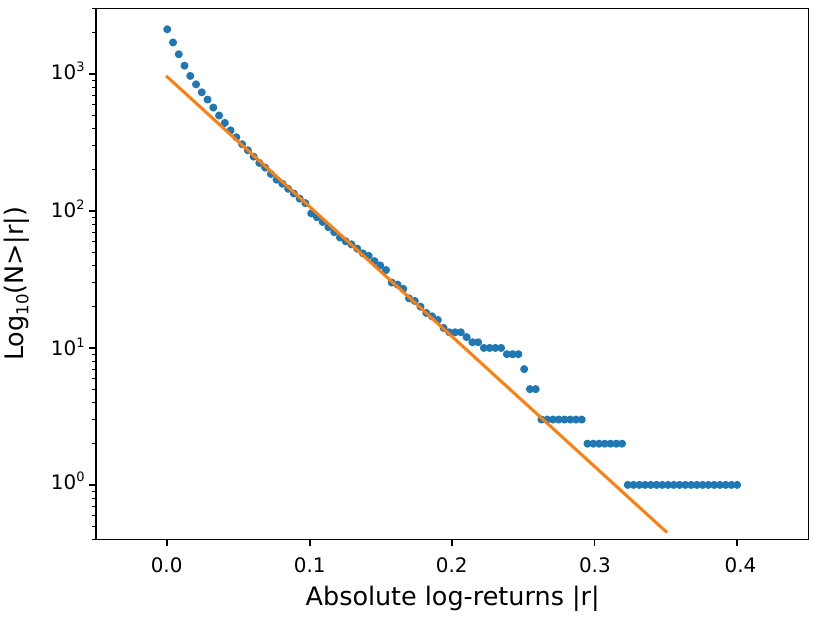}
			\caption{Number of absolute log-returns equal or higher than a specific value. The dotted curve corresponds to experimental data, whereas the solid orange curve corresponds to fitter data.}
			\label{fig:Richter}
		\end{figure}

		Fig. \ref{fig:Richter} shows that BTC obeys the Gutenberg-Richter law with coefficients $a\sim 3$ and $b\sim 9.5$. The obtained $b$ coefficient is higher than those obtained for earthquakes (typically $\lesssim 2.5$). This high $b$-value can be attributed to a strong swarming of returns. Unlike earthquakes, financial systems are constantly producing returns that correspond financial quakes according to our analogy. Thus, high returns are accompanied by a high number of low returns, which produces the observed high $b$-value.
		
		\subsection{Persistence of prices in the time series of bitcoin}		
			The analogy between financial systems and critical phenomena has been the object of much research. For instance, the concept of persistence, originally developed for spin systems, has been used to analyze stock markets\cite{Constantin2005}. Global persistence is defined as the probability $P_g(t)$ that a global random variable associated with the order parameter (magnetization, for instance) maintains its sign until time $t$\cite{Majundar}. Furthermore, it has been shown that the global persistence exhibits a critical behavior that satisfies a simple finite-size scaling relation:
			
			\begin{equation}
				P_g(t)=t^{-\theta_g}f(t/L^z),
			\end{equation}
			where $z$ is a dynamical exponent, and $\theta_g$ is the global persistent exponent for a system of size $L$. In the thermodynamical limit where $L\rightarrow\infty$, a power law is expected for this probability at a critical temperature $T=T_C$:
			
			\begin{equation}
				P_g(t)\sim t^{-\theta_g}.
			\end{equation}
			A deviation of this critical exponent is observed away from the critical temperature.
			The critical exponent $\theta_g$ has been calculated for a myriad of systems, including spin models\cite{Schulke1997,Silva2003} and spatial games\cite{Silva2006}.
			
			For stock markets, the persistence is given as the probability that the price of an asset is greater than or equal to its initial value until an instant $t$, i.e. $P_+(t)=\text{Pr}\left\{p(t')\geq p(0), t'=0\hdots t\right\}$. Alternatively, we can also define a negative version for the persistence given by $P_-(t)=\text{Pr}\left\{p(t')\leq p(0),t'\hdots t\right\}$. The global persistence is then calculated as the mean of both branches: $P_g(t)=\nicefrac{1}{2}\left[P_+(t)+P_-(t)\right]$.
			
			Although the stock market has been shown to produce robust power laws for the persistence\cite{Constantin2005}, the economic crisis of 2008 offered a natural laboratory to test its concept. Persistence has been studied in this situation and its exponent was successfully used to characterize a critical phenomenon\cite{Silva2010}.
			
\begin{figure}
				\centering
				\includegraphics{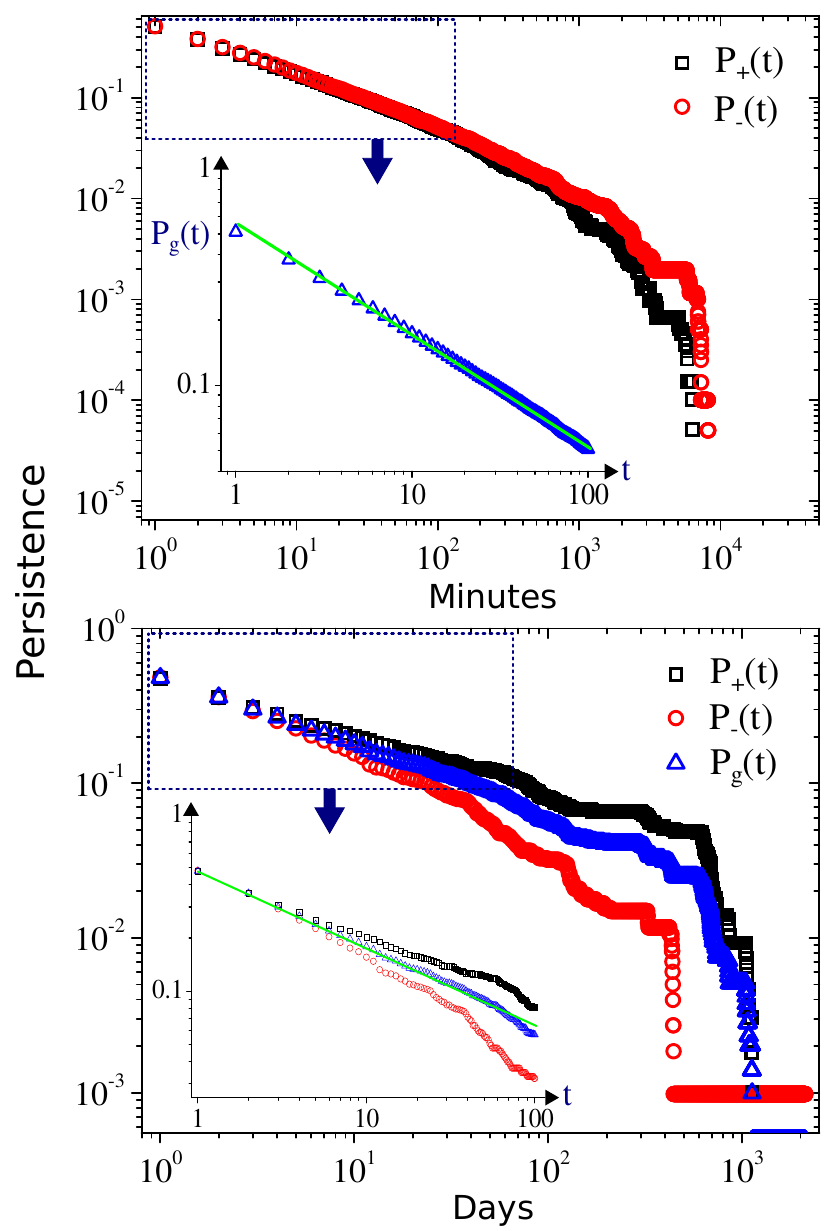}
				\caption{Top) Persistence for 1-minute prices. The inset shows $P_g(t)$ (blue triangles) and a linear fit for the first 100 minutes. Bottom) Persistence for daily prices. The inset shows $P_g(t)$ (blue triangles) as a bisector between $P_+(t)$ (red circles) and $P_-(t)$ (black squares) for the first 100 days. A power law behavior is clear in both cases.}
				\label{fig:Pers}
			\end{figure}

			Given the distinct nature of crypto assets, it is natural to ponder whether bitcoin produces similar results. In order to answer that an alternative algorithm was devised to account for the limited BTC time series. A histogram was constructed with the periods the prices stay above ($P_+$) or below ($P_-$) an initial value. This was conducted for $4\times 10^4$ different initial values randomly sorted. The persistence is finally computed from its complementary cumulative distribution as shown in Fig. \ref{fig:Pers}.
			
			We found a persistence exponent $\theta_g=0.543(4)$ for 1-minute prices, and $\theta_g=0.471(5)$ for daily prices. These values are near the value obtained for the international stock market ($\theta_g\approx 0.5\pm0.02$)\cite{Constantin2005}, and for the Brazilian stock market ($\theta_g=0.568(1)$)\cite{Silva2010}. This suggests that both the stock and bitcoin markets share similar statistical mechanisms.

	\section{Conclusion}
	We have studied bitcoin as a digital financial asset that is devoid of a central (un)coordinating agent. Along this paper we have checked a set of stylized facts for bitcoin that are commonly found in standard financial assets. We have found that: \emph{i}) the logarithmic returns of bitcoin exhibit fat tails, \emph{ii}) bitcoin returns show aggregational Gaussianity, \emph{iii}) BTC returns exhibit fluctuation scaling, \emph{iv}) its volatility tends to cluster in time, \emph{v}) the correlation between the volume and volatility for BTC is always positive, and \emph{vi}) long range returns predict the fine structure better than the other way around. Moreover, we presented an analogy between the volatility and natural occurring quakes and found that BTC obeys both a generalized Omori law and a Gutenberg-Richter law. Finally, we presented results about the persistence of bitcoin prices and showed that: \emph{vii}) BTC shows persistence with power law exponent $\theta_g\approx 0.5$ as found in standard financial markets.
	
	\section*{References}
	\nocite{*}
	\bibliography{paper}
	
\end{document}